\documentclass{epl}
\usepackage{amssymb,graphicx}
\usepackage{times}
\newcommand{\fig}[2]{\includegraphics[width=#1]{./figures/#2}}

\newlength{\bilderlength}

\newlength{\figsize}
\newcommand{\sgn}{{\mathrm{sgn}}}
\newcommand{\rme}{{\mathrm{e}}}
\newcommand{\rmd}{{\mathrm{d}}}

\title{How to measure Functional RG fixed-point 
functions for dynamics and at depinning}
\shorttitle{Functional RG fixed-point functions for dynamics...}
\author{Pierre Le Doussal and Kay J\"org
Wiese}
\shortauthor{P.\ Le Doussal and K.J.\ Wiese}
 \institute{CNRS-Laboratoire de Physique
Th{\'e}orique de l'Ecole Normale Sup{\'e}rieure, 24 rue Lhomond,
75231 Paris Cedex, France, and KITP, UCSB, Santa Barbara, CA 93106-4030, USA.}

\date{\small\today}

\begin{document}

\maketitle
\begin{abstract}
We show how the renormalized force correlator $\Delta(u)$, the
function computed in the functional RG (FRG) field theory, can be
measured directly  in numerics and experiments on the {\it dynamics}
of elastic manifolds in presence of pinning disorder. For
equilibrium dynamics we recover the relation obtained recently in
the statics between $\Delta(u)$ and a physical observable. Its
extension to depinning reveals interesting relations to stick-slip
models of avalanches used in dry friction and earthquake dynamics.
The particle limit ($d=0$) is solved for illustration: $\Delta(u)$
exhibits a cusp and differs from the statics. We propose that
the FRG functions be measured in wetting and magnetic
interfaces experiments.

\end{abstract}

Models involving elastic objects driven through random media are
important for numerous physical systems and phenomena including
magnets \cite{domain-walls-exp},
superconductors\cite{review_pinning_russes.review_nattermann},
density waves \cite{cdwbrazovskii03}, wetting \cite{rolley}, dry
friction \cite{charlaix}, dislocation and crack propagation
\cite{ebouchaud}, and earthquake dynamics \cite{earthquake}. There
has been progress in qualitative understanding of, e.g.\ the
existence of a depinning treshold for persistent motion at zero
temperature $T=0$, scale invariance at the threshold and the analogy
with critical phenomena, collective pinning and roughness exponents,
avalanche motion at $T=0$, and ultra-slow thermally activated creep
motion over diverging barriers. These phenomena are predicted by
theory, i.e.\ phenomenological arguments
\cite{review_pinning_russes.review_nattermann}, mean field models
\cite{fisher85}, functional renormalisation group
\cite{FRG,narayan-fisher93,chauve01,ledoussal02}, and were seen in
numerical studies. Experimental evidence for creep motion was found
in vortex lattices, in ferroelectrics, and in magnetic interfaces in
ferromagnets \cite{ferroelectrics.zeldov,domain-walls-exp}. Some
cases exhibit clear discrepancies with the simplest theories, e.g.\
the depinning of the contact line of a fluid
\cite{rolley,contactkpz}. Even when agreement exists, much remains
to
be done for a precise comparison. 

Recent theoretical progress makes these quantitative tests possible.
For interfaces, powerful algorithms now allow to find the exact
depinning threshold and critical configuration on a cylinder
\cite{rosso2003} and to study creep dynamics \cite{kolton06}. The
functional RG has been extended beyond the lowest order (one loop),
and it was shown that differences between statics and depinning
become manifest only at two loops \cite{chauve01,ledoussal02}. The
FRG is the candidate for a field theoretic description of statics
and depinning, beyond mean field. It captures the complex glassy
physics of numerous metastable states at the expense of introducing,
rather than a single coupling as in standard critical phenomena, a
function, $\Delta(u)$, of the displacement field $u$, which flows to
a fixed point (FP) $\Delta^*(u)$. This FP is non-analytic, as is the
effective action of the theory. At a qualitative level, $\Delta(u)$
can be interpreted as the coarse-grained correlator of the random
pinning force and its cusp singularity at the FP,
$\Delta'(0^+)=-\Delta'(0^-)$, is related to shock singularities in
the coarse grained force landscape, reponsible for pinning. Until
now however, comparison between experiments, numerics and FRG was
mostly about critical exponents.

The aim of this paper is to make precise statements concerning the
physics of {\it dynamical} FRG and propose experimental and
numerical tests. Recently a relation was found \cite{pld} between
the FRG coupling functions $\Delta(u)=-R''(u)$ and {\it
observables}, suggesting a method to measure these functions {\it in
the statics}. The idea is to add to the disorder a parabolic
potential (i.e., a mass $m$) with a variable minimum location $w$.
The resulting sample-dependent free energy $\hat V(w)$ defines a
renormalized random potential whose second cumulant
is proved to be {\it the same} $R(w)$ function as defined
in the replica field theory  -- deviations arising only in higher
cumulants \cite{pld}. This holds for any internal dimension $d$ of
the elastic manifold, any number of components $N$ of its
displacement field $u(x)$, and any $T$. At $T=0$, the (minimum
energy) configuration $u(x;w)$ is unique and smoothly varying with
$w$, except for a discrete set of shock positions where $u(x;w)$
jumps between degenerate minima. The limit of a single particle in a
random potential ($d=0$) maps to decaying Burgers turbulence, and
the statistics of the shocks can in some cases be obtained, yielding
exact result \cite{pld} for $\Delta(u)$.

This method was used recently \cite{alan} to compute numerically the
zero-temperature FRG fixed-point function $\Delta(u)$ in the
statics, for interfaces ($N=1$), using powerful exact minimization
algorithms. Random bond, random field and periodic disorder were
studied in various dimensions $d=0,1,2,3$. The results were found
close to 1-loop predictions and deviations consistent with 2-loop
FRG. A linear cusp was found in any $d$ and the functional shocks
leading to this cusp were seen. The cross-correlation for two copies
of disorder was also obtained and compared to a recent FRG study of
chaos \cite{chaospld}. The main assumptions and central results of
the FRG for the {\it statics} were thus confirmed.  It is important
to extend these methods to the dynamics of pinned objects and to the
depinning transition.

In this Letter we extend the method of Ref.~\cite{pld} to the
dynamics. Using a slow, time-dependent, harmonic potential we show
how the various terms in the effective  dynamical action identify
with the FRG functions. The $T>0$ equilibrium dynamics reduces to
the same definition as used for the statics.  We describe the
extension to depinning at $T=0$.  There the manifold is pulled by a
quasi-static harmonic force (i.e.\ a spring of strength noted
$m^2$), and we show how the statistics of the resulting jumps
directly yields the critical force and the FRG functions, and how
they converge to fixed forms as $m \to 0$. The model is similar to
some stick-slip models used e.g.\ in dry friction
\cite{charlaix,zapperi1} and earthquake dynamics \cite{earthquake}.
The present method provides a different way to look at these
problems in numerics and experiments, in addition to giving a
precise meaning to quantities computed in the field theory. In
particular we discuss the identification of the critical force, the
statistics of the jumps, using for illustration a graphical
construction in $d=0$. There we compute the FP functions for each
universality class. These exhibit a cusp which we find is rounded by
a finite velocity. These effects could be tested in experiments, as
discussed at the end.

We consider the equation of motion for the overdamped dynamics of an
elastic manifold parameterized by its time-dependent displacement
field $u(x,t)$:
\begin{eqnarray}
&& \eta \partial_t u(x,t) = F_x[u(t);w(t)] \label{eq:eqmo} \\
&& F_x[u;w] = m^2 (w - u(x)) + c \nabla_x^2 u(x) + F(x,u(x)) \nonumber
\end{eqnarray}
where $F_x[u(t);w(t)]$ is the total force exerted on the manifold (we
note $u(t)=\{u(x,t)\}_{x \in \mathbb{R}^d}$ the manifold
configuration, $x$ being its $d$-dimensional internal coordinate);
$\eta$ is the friction coefficient and $c$ the elastic constant. Here
at the bare level, the random pinning force is $F(x,u)=-\partial_u
V(x,u)$ and the random potential $V$ has correlations
$\overline{V(0,x) V(u,x')}= R_0(u) \delta^{(d)}(x-x')$.  We consider
first bare random bond disorder with a short-ranged $R_0(u)$. At
non-zero temperature one adds the thermal noise $\langle \xi(x,t)
\xi(x',t') \rangle = 2 \eta T \delta(t-t') \delta^d(x-x')$.  We have
added a harmonic coupling to an external variable $w(t)$, a given
function of time (in most cases we choose it uniformly increasing in $t$). This is the simplest generalization of the statics, where
$w(t)=w$ is time-independant. It is useful to define the fixed-$w$
energy
\begin{equation}
{\cal H}_{w}[u]= \int \rmd^d x\, \frac{m^2}{2} (u(x)-w)^2 + V(x,u(x))
\end{equation}
associated to the force $F_x[u;w]=-\frac{\delta H_{w}[u]}{\delta
u(x)}$. If $w(t)$ is an increasing function of $t$ the model
represents an elastic manifold ``pulled'' by a spring. Quasi-static
depinning is studied for $dw/dt \to 0^+$.

We first describe qualitatively how to measure the FRG functions and
later justify why the relation is expected to be exact. Consider the observable $w(t) - \langle \bar u(t)
\rangle$, where $\bar u(t) = L^{-d} \int \rmd^d x\, u(x,t)$ is the center
of mass position, and $\langle \dots \rangle$ denotes thermal
averages, i.e.\ the ground state at zero temperature. It is
the shift between the translationally averaged displacement and the
center of the well, i.e.\ the extension of the spring. It is
proportional to the pulling force on the manifold, hence to the
translationally averaged pinning force minus the friction force,
i.e.\ $w(t) - \bar u(t) = m^{-2}( \eta v(t) - \int_x F(x,u(x,t)))$ (if
we use periodic boundary conditions inside the manifold). Of
particular interest are: 
\begin{eqnarray}
&& \overline{ w(t) - \langle \bar u(t) \rangle } = m^{-2} f_{av}(t) \\
&& \overline{ [w(t) - \langle \bar u(t) \rangle][w(t') - \langle \bar
u(t') \rangle] }^c = m^{-4} L^{-d} D_w(t,t')\ , \nonumber
\end{eqnarray}
where connected means w.r.t.\ the double average $\overline{\langle
... \rangle}$. If we consider a function $w(t)$ such that $dw(t)/dt >
0$, one can also write:
$ D_w(t,t')= \Delta_w(w(t),w(t'))$.
As written, the function $\Delta_w$ may in general depend on the
history $w(t)$. However we expect that for fixed $L,m$ and slow
enough $w(t)$, e.g.\ $w(t)=vt$ with $v \to 0^+$,
one has $\Delta_w(w(t),w(t')) \to \Delta(w(t)-w(t'))$.
This function $\Delta(w-w')$, which is independent of the process $w(t)$, is
the one defined in the F.T., as we will justify below.

Let us start with non-zero temperature, $T>0$, and consider a
process $w(t)$ so slow that the system (with a finite number of
degrees of freedom $(L/a)^d$) remains in equilibrium. In practice it
means that $\dot{w} t_L \ll u(L) $ where $t_L$ is the largest
relaxation time of the system, and $u(L)$ its width. The above
definition is then consistent with the one from the statics, where
it was shown that one can measure the equilibrium free energy in a
harmonic well with fixed $w$ (or its generalization to an arbitrary
$w(x)$), defined through $e^{- \hat V(w)/T} = \int {\cal D}[u]\,
\rme^{ - {\cal H}_w[u]/T }$, and extract from it the pinning energy
correlator $R(w)$. This can be done by measuring the second cumulant
\cite{foot2} $\overline{\hat V(w) \hat V(w')}^c=\hat R[w-w']$, with
$\hat R[w]=L^d \hat R(w)$ for a uniform parabola $w(x)=w$, and using
that $\hat R=R$ \cite{pld}. One equivalently obtains the force
correlator $\Delta(w)$ via the equilibrium fluctuations of the
center of mass $\langle \bar u \rangle_w$ at fixed $w$, i.e.\
$\overline{ (w - \langle \bar u \rangle_w)(w' - \langle \bar u
\rangle_{w'}) }^c = m^{-4} L^{-d} \Delta(w-w')$. In the statics it
is easy to show that $\Delta(w) =- R''(w)$. The potentiality of this
function breaks down in the driven dynamics, or at depinning, as
discussed below.

Let us note at this stage that a second definition can be given using
two ``copies''. Consider two evolutions $u(x;w_1)$ and $u(x;w_2)$ driven
by two (slow) processes $w_1(t)=w_2(t)+w$ of fixed separation, in the
same disorder sample. Then define
\begin{equation}
\overline{ (w_1(t) - \langle \bar u_1(t) \rangle)(w_2(t) - \langle
\bar u_2(t) \rangle) }^c = m^{-4} L^{-d} \Delta_t(w) \label{2copy}
\end{equation}
which is now an equal-time correlation. For a slow equilibrated
motion at $T>0$, it identifies with the static definition. The
general case is discussed below and in \cite{us_long}.

Let us now describe $T=0$ depinning, and restrict to $N=1$. Quasi-static depinning is
studied as the limiting case where $dw/dt \to 0^+$.
One starts in a
metastable state $u_0(x)$ for a given $w=w_0$, i.e.\ a zero-force
state $F_{x}(u_0(x);w)=0$ which is a local minimum of
$H_{w_0}[u]$ with a positive barrier. One then increases $w$. For smooth short-scale
disorder, the resulting deformation of $u(x)$ is smooth. At $w=w_1$, the barrier vanishes.
For $w=w_1^+$ the manifold  moves downward in energy until it is blocked again in
a metastable state $u_1(x)$ which again is  a local minimum of $H_{w_1}[u]$.
We are interested in the center of mass (i.e.\ translationnally
averaged) displacement $\bar u = L^{-d} \int \rmd^d x\, u(x)$. The above
process defines a function $\bar u(w)$ which exhibits jumps at the
set $w_i$. Note that time has disappeared: evolution is only used to find the next
location. The first two cumulants
\begin{equation}
\overline{w - u(w)} = m^{-2} f_c  \label{def-fc} \quad , \quad
\overline{(w - u(w))(w' - u(w')}^c = m^{-4} L^{-d} \Delta(w-w')
\label{def-Delta}
\end{equation}
allow a direct determination (and definition) of the averaged
($m$-dependent) critical force $f_c$ and of $\Delta(w)$, in analogy
to the statics. Note that $u(w)$ depends a priori on the initial
condition and on its orbit but at fixed $m$ one expects an averaging
effect when $w$ is moved over a large region. This is further
discussed below. Note that the definition of the (finite-size)
critical force is very delicate in the thermodynamic limit
\cite{fedorenkofc}. Here the quadratic well provides a clear way to
obtain a stationary state.

\begin{figure}[t]
\centerline{\fig{9cm}{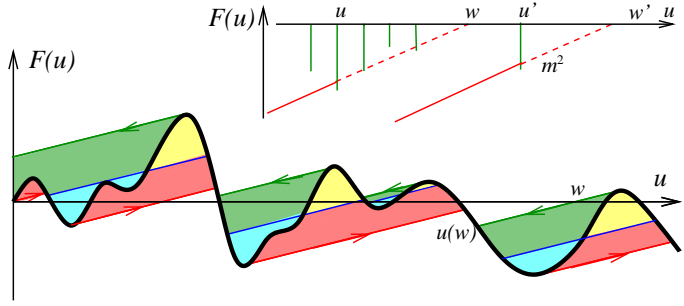}}
\caption{Main plot: Construction of $u(w)$ in $d=0$.  The pinning
force $F(u)$ (bold black line). The two quasi-static motions driven
to the right and to the left are indicated by red and green arrows,
and exhibit jumps (''dynamical shocks''). The position of the shocks
in the statics is shown, for comparison, from the Maxwell
construction (equivalence of light blue and yellow areas, both
bright in black and white). The critical force is $2/L$ times the
area bounded by the hull of the construction. Inset: The same
construction for the forward motion of the discretized model.}
\label{figgraphical}
\end{figure}%
Elastic systems driven by a spring and stick-slip type motion were
studied before, e.g.\ in the context of dry friction. The force
fluctuations, and jump distribution were studied numerically for a
string driven in a random potential \cite{zapperi1}. However, the
precise connection to quantities defined and computed in the field
theory has to our knowledge not been made. The dependence on $m$ for
small $m$ predicted by FRG, $\Delta(w) = m^{\epsilon-2 \zeta} \tilde
\Delta(w m^{-\zeta})$ is consistent with observations of
\cite{zapperi1} but the resulting $\tilde \Delta(w)$ has not been
measured.  Fully connected mean-field models of depinning also
reduce to a particle pulled by a spring, together with a
self-consistency condition, around which one can expand
\cite{narayan-fisher93}. As discussed below, our main remarks here
are much more general, independent of any approximation scheme, and
provide a rather simple and transparent way to attack the problem.

For the qualitative discussion it is useful to study the
model in $d=0$, i.e.\ a particle with equation of motion
\begin{equation}\label{d1}
\eta \partial_{t} u = m^{2} (w-u) + F (u)\ .
\end{equation}
In the quasi-static limit where $w$ is increased slower
than any other time-scale in the problem, the zero
force condition $F (u) = m^{2} (u-w)$ determines $u(w)$ for
each $w$. The graphical construction
of $u(w)$ is well known from studies of dry friction
\cite{charlaix}. When there are several roots one must follow
the root as indicated in Fig.\ \ref{figgraphical}, where
$F (u)$ is plotted versus $m^{2} (u-w)$. This results in
jumps and a different path for motion to the right and to the left.
Let us call $A$ the area of this hysteresis loop (the area of all
colored/shaded regions in Fig.\ref{figgraphical}). It is the
total work of the friction force when moving the center of
the harmonic well quasi-statically once forth and back, i.e.\ the
total dissipated energy. The above definition of the
averaged critical force (\ref{def-fc}), assuming the
landscape statistics to be translationally invariant
(hence replacing disorder averages by translational
ones over a large width $M$) gives
\begin{equation}\label{d4}
f_{c} = m^{2}\,  \overline{( w-u_{w}) }^{\mathrm{tr}} = \frac{m^{2}}{M} \int_{0}^{M}
\rmd w\, \left(w-u_{w} \right) =  \frac{A}{2M}\ ,
\end{equation}
where we have used $\int u \, \rmd w = \int w\, \rmd u$.  One can
check that for $m \to 0$ this definition of $f_c$ becomes identical to
the one on a cylinder, $f_d$, which for a particle ($d=0$) is $2 f_d =
f_d^+-f_d^-=\max_{u} F(u)-\min_{u} F(u)$ with $2 f_d M = \lim_{m \to 0} A(m)$.
(Since $A$ depends on the starting point, this definition holds after a second tour, where the
maximum (minimal) pinning force was selected). Finally, one can compare
with the definition of shocks in the statics. There, the effective
potential is a continuous function of $w$. Therefore, when making a
jump, the integral over the force must be zero, which amounts to the
Maxwell-construction of figure \ref{figgraphical}.

One can compute $f_c$ and $\Delta(w)$ in $d=0$ for a  discrete force
landscape, $F_i$, independently distributed with $P(F)$, and $i$
integer. $u(w)$ is then integer and defined in the inset of figure
\ref{figgraphical}. The process admits a continuum limit for small
$m$, which depends on the behaviour of $P(F)$ in its tails (negative
tail for forward motion). One obtains \cite{us_long} the
distribution of $u(w)$, $P_w(u) \rmd u = \rme^{- a_{w}(u)} \rmd
a_{w}(u)$ where $a_w'(u) = \int_{-\infty}^{m^2(u-w)} P(f) \rmd f$
and $a_{w} (-\infty)= 0$. One also obtains the joint distribution of
$(u(w),u(w'))$, $P_{w;w'} (u,u') = (a'_{w} (u)- a'_{w'} (u)) a'_{w'}
(u') \rme^{-a_{w} (u)- a_{w'} (u')+a_{w'} (u)} \theta(u'-u) +
\delta(u'-u) a'_{w'} (u) \rme^{-a_{w} (u)}$ for $w>w'$.  Define
$\Delta(w)=:m^4 \rho_m^2 \tilde \Delta(w/\rho_m)$ and $f_c =: f_c^0
+ c m^2 \rho_m$. This yields two main classes of universal behaviour
at small $m$. The first contains (i) exponential-like distributions
with unbounded support i.e.\ $\ln P(f) \approx_{f \to - \infty} - A
(-f)^\gamma$ (for which $f_c^0= ((\ln
m^{-2})/A)^{\frac{1}{\gamma}}$) and (ii) distributions with
exponential behaviour near an edge $P(f) \sim e^{- A (f+f_0)^\gamma}
\theta(f+f_0)$ (with $\gamma<0$ and $f_c^0=f_0 - ((\ln
m^{-2})/A)^{\frac{1}{\gamma}})$. For both (i) and (ii) the FP
function is $\tilde \Delta(x)
=\frac{x^2}{2}+\mbox{Li}_2(1-e^x)+\frac{\pi ^2}{6}$ and
$\rho_m=\rho_m^\gamma:=1/(|\gamma| A^{1/\gamma} m^2 (\ln
m^{-2})^{1-\frac{1}{\gamma}})$, $c = \gamma_E$ the Euler constant.
The first class has $\zeta=2$ up to log-corrections. The second
class contains power-law distributions near an edge $P(f)= A
\alpha(\alpha-1)(f+f_0)^{\alpha-2} \theta(f+f_0)$, $\alpha>1$, for
which $c= - \Gamma(1 + \frac{1}{\alpha})$, $m^2
\rho_m=(m^2/A)^{\frac{1}{\alpha}}$ and $f_c^0=f_0$. The FP depends
continuously on $\alpha$ with $\tilde
\Delta(w)=-\Gamma(1+\frac{1}{\gamma})
\Gamma(1+\frac{1}{\gamma},w^\gamma) + w \Gamma(1+\frac{1}{\gamma})
e^{-w^\gamma} + \int_0^\infty \rmd y e^{-(y+w)^\gamma + y^\gamma}
\Gamma(1+\frac{1}{\gamma},y^\gamma)$ where $\Gamma(a,x) =
\int_{x}^\infty \rmd z z^{a-1} e^{-z}$; it has $\zeta=2-2/\alpha$
\cite{footnoteII}. Hence, despite the fact that $d=0$ is dominated
by extreme statistics (e.g. the distribution of $\rho^{-1}_m
(w-u(w))$ converges to the Gumbel and Weibul
distributions for class I and II respectively) it still exhibits
some universality in cumulants and in all classes $\Delta(u)$ has a
cusp non-analyticity at $u=0$. We have checked the above scaling
functions {\em and} amplitudes numerically, with excellent
agreement.

We now come back to the interface, $d>0$, and note that the manifold
in the harmonic well can be approximated by $(L/L_m)^d$ roughly
independent pieces with $L_m \sim 1/m$. The motion of each piece
resembles the one of a particle, i.e.\ a $d=0$
model, but with a rescaled unit of distance in the $u$ direction,
$u_m \sim L_m^\zeta \sim m^{- \zeta}$.  The ``effective-force''
landscape seen by each piece becomes uncorrelated on such distances,
and its amplitude scales as $F_m \sim m^2 u_m$. Hence one is in a
bulk regime not dominated by extremes, i.e.\ $\Delta(w)$ probes only
motion over about one unit. It is easy to check on Fig.~1 that an
arbitrary initial condition joins the common unique orbit after
about one correlation length. Hence the $d=0$ model suggests that
starting the quasi-static motion in $u_0$ and driving the manifold
over $w \sim L_m^\zeta$ should then result in all orbits converging.
Hence the definitions (\ref{2copy}) and (\ref{def-Delta}) are
equivalent for $N=1$. An interesting crossover to $d=0$ behaviour
and extremal statistics occurs if $L < L_m$.

\begin{figure}[t]
\centerline{{\setlength{\unitlength}{1mm}
\fboxsep0mm
\mbox{\begin{picture}(86,50)
\put(0,0){\fig{86mm}{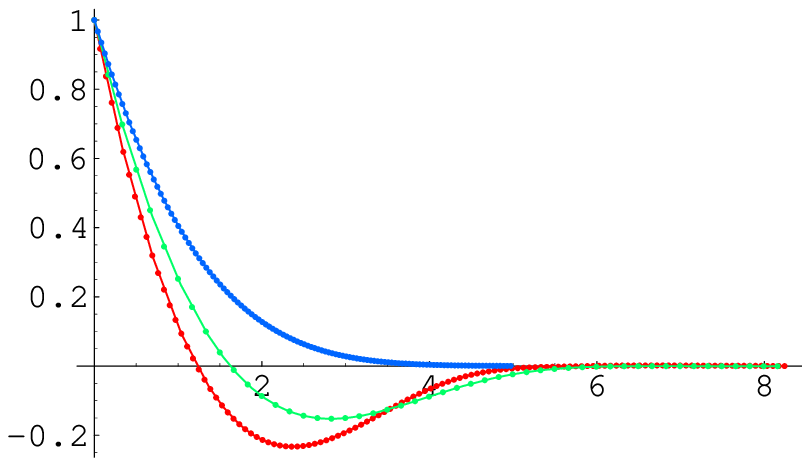}}
\put(0,48){$\Delta (u)$}
\put(84,8){$u$}
\put(13,5){$m^{2}=1$}
\put(40,14){$m^{2}=0.003$}
\put(27,48){$\Delta' (u)$}
\put(84,47){$u$}
\put(32,18){\mbox{\fig{55mm}{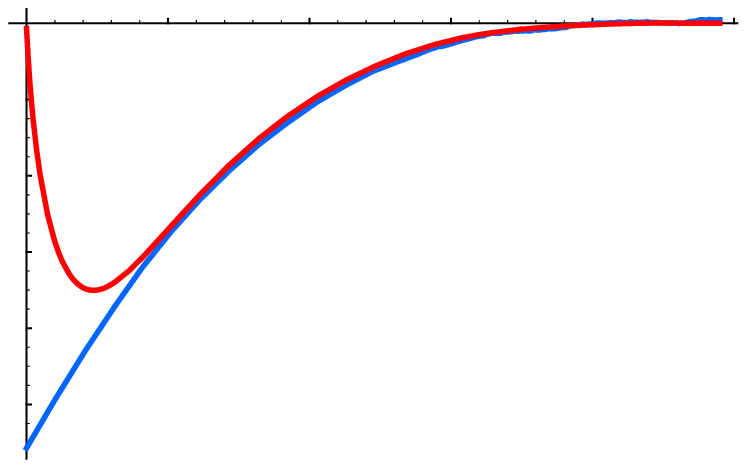}}}
\end{picture}}}}
\caption{(a): Main plot. The measured $\Delta (u)$ for RB disorder
distributed uniformly in $[0,1]$, and rescaled such that $\Delta
(0)=1$ and $\int_{0}^{\infty} \rmd u |\Delta (u)|=1.$ From bottom
(which has $\int_{0}^{\infty} \rmd u \Delta (u)\approx 0$) to top
the mass decreases from
$m^{2} =1$ to $
0.5$ to $0.003$. One observes a clear
crossover from
RB to RF. \newline
(b): Inset. $\Delta' (u)$ for a driven interface at vanishing (blue,
$\Delta' (0)<0$) and finite velocity (red, $\Delta' (0)=0$).}
\label{f:RBRF-finvel}
\end{figure}%
Note that the averaged critical force, defined in (\ref{def-fc}),
should, for $d>0$, go to a finite limit, with $f_c(m)=f_c^\infty + B
m^{2-\zeta}$ from finite size scaling. Although $f_c$ is not universal
and depends on short-scale details, one easily sees that $-m
\partial_m f_c(m)$ depends only on one unknown scale. We note that the
definition (\ref{def-fc}) coincides with the one proposed recently
as the maximum depinning force for all configurations having the
same center of mass $u_0$ \cite{fedorenkofc}. Since  $\bar u - w$ is
a fluctuating variable of order $(L/L_m)^{-d/2}$, the definition is
the same as the above in the limit where $L\to\infty$, before $m\to
0$. The single $w$ distribution  is obtained from the distribution
of $w - u(w)$ if all modes have a mass.

Measurements of $\Delta(w)$ reveal interesting features in any $d$.
At the bare level, the disorder of the system is of random-bond type
(i.e.\ potential).  As the mass is decreased, one should observe a
crossover from random-bond to random-field disorder. Also a finite
velocity should round the cusp singularity. These features are well
visible in the $d=0$ toy model as illustrated in
Fig.~\ref{f:RBRF-finvel}a (quasi-static evolution in a model with
$F_i=V_{i+1}-V_i$ and $V_i$ uncorrelated) and Fig.\
\ref{f:RBRF-finvel}b (Langevin dynamics at finite $v$) and can be
obtained analytically in that case \cite{us_long}.

We now sketch the exact relation to the FT definitions. From now
on we use condensed notations $u(x,t)=u_{xt}$ and so on. The bare MSR action
functional can be parameterized as ${\cal S}[u,\hat u] = \hat u \cdot g^{-1} \cdot u + \hat u \cdot A^{(0)}[u]
- \frac{1}{2} \hat u \cdot B^{(0)}[u] \cdot \hat u + O(\hat u^3)$ with
$g_{xy}$ an arbitrary (time independent) symmetric matrix (standard
choice being $g^{-1}_q = q^2 + m^2$), $A^{(0)}[u]_{xt} = \eta
\partial_{t} u_{xt}$ and $B^{(0)}[u]_{xt,x't'} = 2 \eta T \delta_{xx'}
\delta_{tt'} + \Delta_0(u_{xt}-u_{x't'}) \delta_{xx'}$. We denote $u
\cdot v := \int_{xt} u_{xt} v_{xt}$ (and additional index
contraction for $N>1$), $A^{(0)}$ and $B^{(0)}$ are respectively
vector and matrix functionals. The effective action $\Gamma[u]$ can
be parameterized identically with $A^{(0)}[u] \to A[u]$ and
$B^{(0)}[u] \to B[u]$. It is obtained from the generating function:
$W[w,\hat w] = \ln \int {\cal D}[u]\, {\cal D}[\hat u]\, \exp(
-{\cal S}[u,\hat u]+ \hat u \cdot g^{-1} \cdot w + \hat w \cdot
g^{-1} \cdot u )$ through a Legendre transform: $W[w,\hat w] +
\Gamma[u,\hat u] = \hat u \cdot g^{-1} \cdot w + \hat w \cdot g^{-1}
\cdot u$. It can be expanded\cite{foot2} as: $W[w,\hat w] = \hat w
\cdot g^{-1} \cdot w - \hat w \cdot \hat A[w] + \frac{1}{2} \hat w
\cdot \hat B[w] \cdot \hat w + O(\hat w^3)$. This functional
directly generates correlations (\ref{def-fc}) and
(\ref{def-Delta}), in a more general form: $\overline{w_{xt}-\langle
u_{xt} \rangle_w } = g_{xy} \hat A_{yt}[w]$ and
$\overline{(w_{xt}-\langle u_{xt} \rangle_w)(w_{x't'}- \langle
u_{x't'} \rangle_w ) }^c = g_{xy} g_{x'y'} \hat B_{yt,y't'}[w]$.
Hence $\hat A$ and $\hat B$ are observables which can be measured,
i.e.\ for a uniform $w_{xt}=w_t$, $f_{av}(t)=\frac{1}{L^d} \int_y
\hat A_{yt}[w]$ and $D_{w}(t,t')= \frac{1}{L^{d}} \int_{yy'} \hat
B_{yt,y't'}[w]$, which for slow $w_t$ should go to $f_c$ and
$\Delta(w_t - w_{t'})$ respectively. The question is how to relate
them to the effective action, i.e.\ generalize the relation $\hat
R=R$ from the statics. To this aim we perform a Legendre transform.
Details are given in \cite{us_long}. The result is $A[u[w]]
=\hat{A}[w]$, (where we have defined $u[w]:=w-g \cdot \hat A[w]$,
i.e. $u_{xt}[w]:=\langle u_{xt} \rangle_w$) and $\hat{B}[w] = (\rmd
\hat{u}/\rmd \hat{w})^{t}\cdot B[u[w]] \cdot \rmd \hat{u}/\rmd \hat{w}$ with
$\rmd \hat{u}/\rmd \hat{w}= 1- (\nabla_{w} \hat A [w])^t \cdot g$.
Now consider $w$ uniform in space $w_{xt}=w_t$. Then $\hat
A_{yt}[w]$ is $y$-independent. In the limit of infinitely slow
monotonous $w_t$ one expects $A_{yt}[w] \to f_c \,\sgn( \dot w)$ hence
$u_{xt}[w] \to w_t+ \sgn( \dot w) f_c/m^2$ (with usually $f_c=0$ at
$T>0$). From STS it implies $A[w]=\hat A[w]$ in that limit for
uniform $w$. Similarly one finds $D_w(t,t')=\sum_{xy} B_{xt,yt'}[w]$
for infinitely separated times (with fixed $w_t-w_t'$) since $\rmd
\hat{u}/\rmd \hat{w}=g \frac{\delta u[w]}{\delta w} \to 1$, i.e. the
disorder-averaged response function becomes trivial at zero frequency
(due again to STS). Hence the measured $\Delta(w)$ in
(\ref{def-Delta}) is -- to all orders -- the one defined in the FT.

We have generalized \cite{us_long} the above method to a manifold
driven in $N$ dimensions (e.g.\ a flux line in a 3D superconductor).
For a particle, and fixed $m$, we have seen numerically that
different initial conditions converge. The Middleton theorem no
longer holds, and particles can pass by each other. To probe
transverse motion and correlations $\Delta(w)$ in the transverse
direction, we use the two-copy definition (\ref{2copy}). Finally
thermal rounding of depinning, creep and crossover from statics to
depinning can be studied more precisely by this method.

To conclude let us propose that $\Delta(w)$ be measured directly in
experiments, which would represent an important test of the theory
and the underlying assumptions. Creep and depinning of magnetic
domains in thin films with surface step disorder have been
investigated using imaging \cite{domain-walls-exp}: adding a
magnetic field gradient should allow to confine the interface in an
effective quadratic well, whose strength and position can be varied
(hence probing both statics and dynamics). In contact lines of
fluids it is capillarity and gravity which provide the quadratic
well, and provided large scale inhomogeneities can be controlled,
$\Delta(u)$ could be measured from statistics on lengths larger than
the capillary length (i.e. $L_m$ here). We hope this will stimulate
further numerical \cite{alberto} and experimental \cite{rolley2}
studies.

We thank A. Fedorenko, E. Rolley, A. Rosso and S. Moulinet for
useful discussions, the KITP for hospitality and support from ANR
(05- BLAN-0099-01).


\begin{thebibliography}{100}
\bibitem{foot2} STS implies that $\overline{\hat V(w)}$ is a constant and that $\hat A[w]$ and $\hat B[w]$
are invariant under the change $w_{xt} \to w_{xt} + \phi_x$, and the
same for the  functionals  $A$, $B$, $A^0$, and $B^0$.
\bibitem{footnoteII}
Distributions with unbounded tails $p(f) \sim A \alpha (\alpha+1)
f^{2+\alpha}$ at large negative $f$ correspond to $\alpha \to -
\alpha$, i.e.\ $\zeta=2+2/\alpha$, $f_c^0= 0$, $m^2
\rho_m=(A/m^2)^{1/\alpha}$, $c=\Gamma(1 - \frac{1}{\alpha})$ and
Frechet class of single $w$ statistics. They are dominated by rare
events with no finite continuum limit for cumulants of order larger
than $\alpha$.
\end{thebibliography}
\end{document}